# Protein structure and sequence re-analysis of 2019-nCoV genome does not indicate snakes as its intermediate host or the unique similarity between its spike protein insertions and HIV-1

Chengxin Zhang, Wei Zheng, Xiaoqiang Huang, Eric W. Bell, Xiaogen Zhou, Yang Zhang*

*Department of Computational Medicine and Bioinformatics, University of Michigan, Ann Arbor, MI 48109, USA*

*Correspondence should be addressed to Dr. Yang Zhang (Email: zhng@umich.edu)

## Abstract

As the infection of 2019-nCoV coronavirus is quickly developing into a global pneumonia epidemic, careful analysis of its transmission and cellular mechanisms is sorely needed. In this report, we re-analyzed the computational approaches and findings presented in two recent manuscripts by Ji *et al*. (https://doi.org/10.1002/jmv.25682) and by Pradhan *et al*. (https://doi.org/10.1101/2020.01.30.927871), which concluded that snakes are the intermediate hosts of 2019-nCoV and that the 2019-nCoV spike protein insertions shared a unique similarity to HIV-1. Results from our re-implementation of the analyses, built on larger-scale datasets using state-of-the-art bioinformatics methods and databases, do not support the conclusions proposed by these manuscripts. Based on our analyses and existing data of coronaviruses, we concluded that the intermediate hosts of 2019-nCoV are more likely to be mammals and birds than snakes, and that the "novel insertions" observed in the spike protein are naturally evolved from bat coronaviruses.

**Introduction**

The 2019 novel conronavirus, or 2019-nCoV, recently caused a major pneumonia outbreak in Wuhan, China and has spread to at least 20 countries in the world. As of February 7, 2020, there are more than 30,000 patients diagnosed with 2019-nCoV infection in China alone with more than 700 deaths. Understanding the viral infection mechanisms and animal hosts are of high urgency for the control and treatment of the 2019-nCoV virus. While it is now commonly recognized that bats such as *Rhinolophus affinis* may serve as the natural reservoir of 2019-nCoV[1], it is still unclear which animal serves as an intermediate host that brought the bat coronavirus to human hosts. During the 2019-nCoV's infection of host cells, a critical virion protein is the Spike surface glycoprotein, also known as the S protein. Spike proteins constitute the outermost component in a coronavirus virion particle and are responsible for viral recognition of Angiotensin Converting Enzyme 2 (ACE2), a transmembrane receptor on mammalian hosts that facilitate viral entrance into host cells[1-2]. Therefore, the spike protein largely determines host specificity and infectivity of a coronavirus. In this report, we re-analyzed two recent studies[3-4], which have spurred numerous interests and discussions in the community and the society, regarding the sequence and structure of spike protein in the 2019-nCoV and the identification of its intermediate hosts.

# Results

## 2019-nCoV Spike Protein does not Include Insertions Unique to HIV-1

In a recent manuscript entitled "Uncanny similarity of unique inserts in the 2019-nCoV spike protein to HIV-1 gp120 and Gag" [3], Pradhan *et al*. presented a discovery of four novel insertions unique to 2019-nCoV spike protein (**Figure 1**). They further concluded that these four insertions are part of the receptor binding site of 2019-nCoV, and that these insertions shared "uncanny similarity" to Human Immunodeficiency Virus 1 (HIV-1) proteins but not to other coronaviruses. These claims have resulted in considerable public panic and controversy in the community. To investigate whether the conclusions by Pradhan *et al*. are scientifically precise, we re-analyzed the structural location and sequence homology of the four spike protein insertions discussed therein.

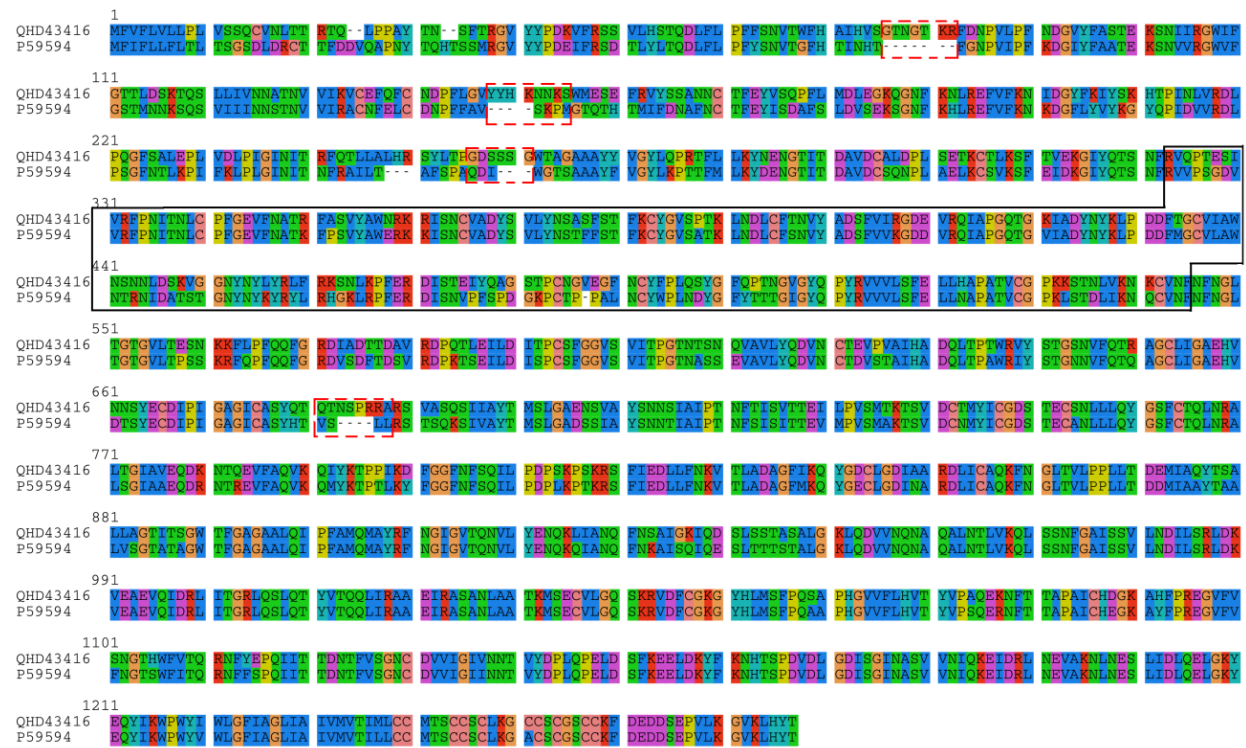

**Figure 1.** Sequence alignment of 2019-nCoV spike protein (NCBI accession: QHD43416) and SARS-CoV spike protein (UniProt ID: P59594). The four "novel" insertions "GTNGTKR" (IS1), "YYHKNNKS" (IS2), "GDSSSG" (IS3) and "QTNSPRRA" (IS4) by Pradhan *et al.* are highlighted in dashed rectangles. We noted that these fragments are not *bona fide* "insertions"; in fact, at least three out of all four fragments are also shared with Bat Coronavirus RaTG13 spike glycoprotein (NCBI accession: QHR63300.1), as shown in **Table 1-3**. Nevertheless, we still refer these fragments as "insertions" in this manuscript for consistency with the original report. The receptor binding domain of spike is marked within the solid box, which corresponds to residue positions 323 to 545 in the above alignment.

Since the structure of the spike protein in 2019-nCoV is not yet available, we used C-I-TASSER[5] to model its tertiary structure, as part of our efforts for full genome structure and function analyses of 2019-nCoV that are available at https://zhanglab.ccmb.med.umich.edu/C-I-TASSER/2019-nCov/. The C-I-TASSER model has a reasonable confidence score which corresponds to an estimated TM-score of 0.61.[6] The 2019-nCoV spike model was then assembled with the human ACE2 structure (PDB ID: 6ACJ)[7] by DEMO[8] to form a spike-ACE2 complex. As shown in **Figure 2**, all four insertions are located outside the Receptor Binding Domain (RBD) of spike, in contrast to the original conclusion made by Pradhan *et al.* which stated that the insertions are located on the interface with ACE2.

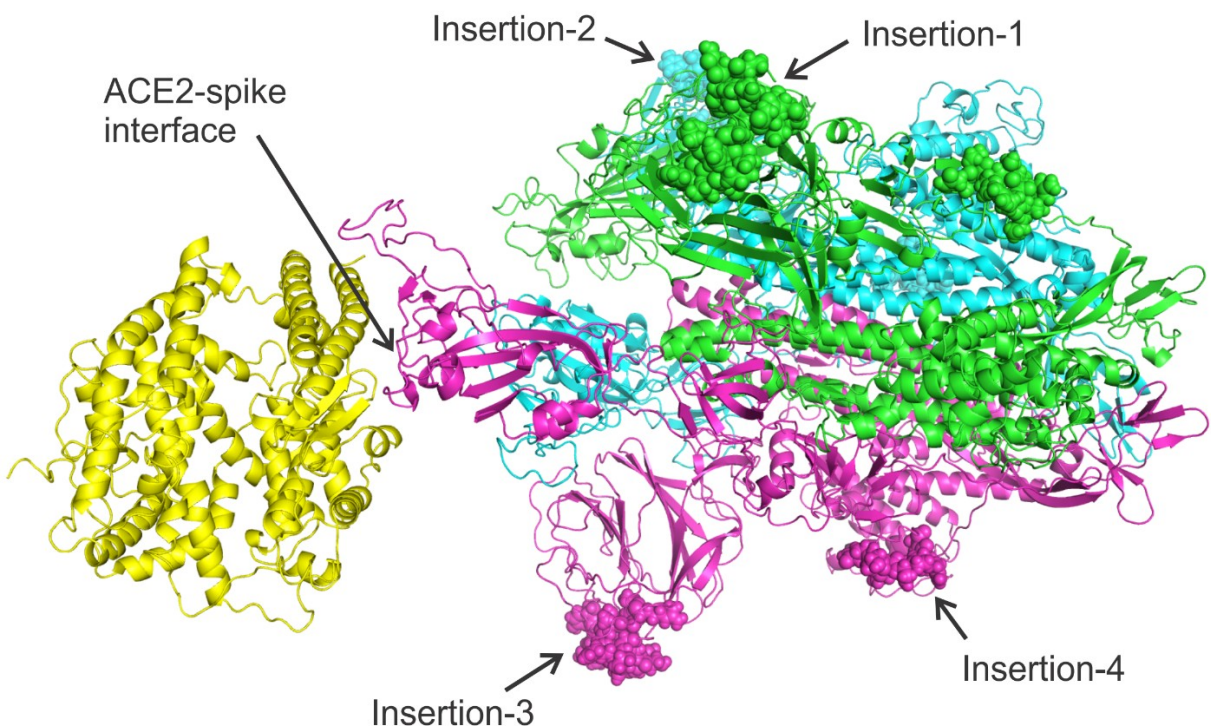

**Figure 2**. Complex structure model between human ACE2 (left yellow) and 2019-nCoV spike protein trimer (right, with three chains colored in magenta, cyan, and blue respectively) constructed by C-I-TASSER. The four insertions are shown as spheres. During different stages of coronavirus infection, the spike proteins may be post-processed (i.e. cleaved) to produce different isoforms. Therefore, the eventual spike complex might not include all residues of a full-length spike protein. Nevertheless, we construct the complex model using full-length spike sequence to illustrate the locations of the four insertions.

To investigate viral homologs of the four insertions, we further performed a BLAST sequence search of these four insertions against the non-redundant (NR) sequence database, restricting the search results to viruses (taxid:10239), but leaving other search parameters at default values. The choice of BLAST instead of the more sensitive PSI-BLAST algorithm[9] is to emulate the original report by Pradhan *et al.* which mainly aimed to identify near identical sequences. The top 5 sequence homologs (including the query itself) identified for each insertion are listed in **Tables 1-4**. In contrast to the previous claim that the four insertions are unique to 2019-nCoV and HIV-1, all four insertion fragments can be found in other viruses. In fact, an HIV-1 protein is among the top BLAST hits for only one of the four insertion fragments, while three of the four insertion fragments are found in bat coronavirus RaTG13. Moreover, partly due to the very short length of these insertions, which range from 6 to 8 amino acids, the E-value of the BLAST hits, which is a parameter used by BLAST for assessing the statistical significance of the alignments and usually needs to be below 0.01 to be considered as significant[9], are all greater than 4, except for a bat coronavirus hit for IS2. These high E-values suggest that the majority of these similarities are likely to be coincidental.

**Table 1.** BLAST search result for IS1. For Table 1 to 4, if there are multiple redundant hits for the same gene from different strains of the same species removed, only one hit is shown. The residue non-identical to query is highlighted in bold. Sequence identity is calculated as the number of identical residues divided by query length. Only the sequence portion aligned to the query is shown. In this table, we also list the closest BLAST hit from bat coronavirus, which is known to be closely related to 2019-nCoV [1].

| NCBI accession | Sequence | E-value | Seq. Identity | Species |
|---|---|---|---|---|
| Query | GTNGTKR | 27 | 1.00 | 2019-nCoV |
| APC94153.1 | GTNGTKR | 28 | 1.00 | uncultured marine virus |
| AFU28737.1 | -TNGTKR | 224 | 0.86 | Human immunodeficiency virus 1 |
| AVE17137.1 | GT**D**GTKR | 224 | 0.86 | Rat astrovirus Rn/S510/Guangzhou |
| QBX18329.1 | -TNGTKR | 224 | 0.86 | *Streptococcus* phage Javan411 |
| QHR63300.1 | GTNG**I**KR | 643 | 0.86 | Bat coronavirus RaTG13 |

**Table 2.** BLAST search result for IS2.

| NCBI accession | Sequence | E-value | Seq. Identity | Species |
|---|---|---|---|---|
| Query | YYHKNNKS | 0.13 | 1.00 | 2019-nCoV |
| QHR63300.1 | YYHKNNKS | 0.13 | 1.00 | Bat coronavirus RaTG13 |
| AUL79732.1 | -YHKNNKS | 4.2 | 0.88 | Tupanvirus deep ocean |
| YP_007007173.1 | YYHK**D**NK- | 8.7 | 0.75 | *Klebsiella* phage vB_KleM_RaK2 |
| ALS03575.1 | YYHKNN-- | 12 | 0.75 | Gokushovirus WZ-2015a |

**Table 3.** BLAST search result for IS3.

| NCBI accession | Sequence | E-value | Seq. Identity | Species |
|---|---|---|---|---|
| Query | GDSSSG | 1004 | 1.00 | 2019-nCoV |
| QAU19544.1 | GDSSSG | 1003 | 1.00 | Orthohepevirus C |
| AYV78550.1 | GDSSSG | 1004 | 1.00 | Edafosvirus sp. |
| QHR63300.1 | GDSSSG | 1004 | 1.00 | Bat coronavirus RaTG13 |
| QDP55596.1 | GDSSSG | 1004 | 1.00 | Prokaryotic dsDNA virus sp. |

**Table 4.** BLAST search result for IS4.

| NCBI accession | Sequence | E-value | Seq. Identity | Species |
|---|---|---|---|---|
| Query | QTNSPRRA | 1.0 | 1.00 | 2019-nCoV |
| YP_009226728.1 | QTNSPRR- | 8.5 | 0.88 | *Staphylococcus* phage SPbeta-like |
| BAF95810.1 | QTNSPRRA | 35 | 1.00 | *Bovine* papillomavirus type 9 |
| ARV85991.1 | **E**TNSPRR- | 106 | 0.75 | Peach associated luteovirus |
| QDH92312.1 | QTN**A**PR**K**A | 142 | 0.75 | *Gordonia* phage Spooky |

Given that 3 out of the 4 insertion fragments are found in the bat coronavirus RaTG13, it is attempting to assume that these "insertions" may be directly inherited from bat coronaviruses. Currently, there are at least 7 known human coronaviruses (2019-nCoV, SARS-CoV, MERS-CoV, HCoV-229E, HCoV-OC43, HCoV-NL63 and HCoV-HKU1), where many of them, including Severe Acute Respiratory Syndrome-related Coronavirus (SARS-CoV) and Middle East Respiratory Syndrome-related Coronavirus (MERS-CoV), were shown to be transmitted from bat[10-14] (**Figure 3A**). To further examine the evolutionary relationship between the 2019-nCoV genome and the bat coronavirus, in comparison with other human coronaviruses, in **Figure 3B** we created by MUSCLE a multiple sequence alignment (MSA) for all the 7 human coronaviruses and the bat coronavirus RaTG13 that is currently known to be the closest relative to 2019-nCoV[14] among known coronaviruses. Among the 4 "insertions" (ISs) of the 2019-nCoV, IS1 has only 1 residue different from the bat coronavirus, and 3 out of 7 residues are identical with MERS-CoV. IS2 and IS3 are all identical to the bat coronavirus. For IS4, although the local sequence alignment by BLAST did not hit the bat coronavirus in **Table 4**, it has a close evolutionary relation with the bat coronavirus in the MSA. In particular, the first 6 residues in the IS4 fragment "QTQTNSPRRA" from 2019-nCoV are identical to the bat CoV, while the last 4 residues, which were absent in the bat coronavirus or SARS-CoV, have at least 50% identity to MERS-CoV and HCoV-HKU1.

Putting these together, we believe that there is a close evolutionary relation between 2019-nCoV and bat coronavirus RaTG13. The four insertions highlighted by Pradhan *et al*. in the spike protein are not unique to 2019-nCoV and HIV-1. In fact, the similarities in the sequence-based alignments built on these very short fragments are statistically insignificant, as assessed by the BLAST E-values, and such similarities are shared in many other viruses including the bat coronavirus. Structurally, these "insertions" are far away from the binding interface of the spike protein with the ACE2 receptor, as shown in **Figure 2**, which are also contradictory with the conclusion made by Pradhan *et al*.

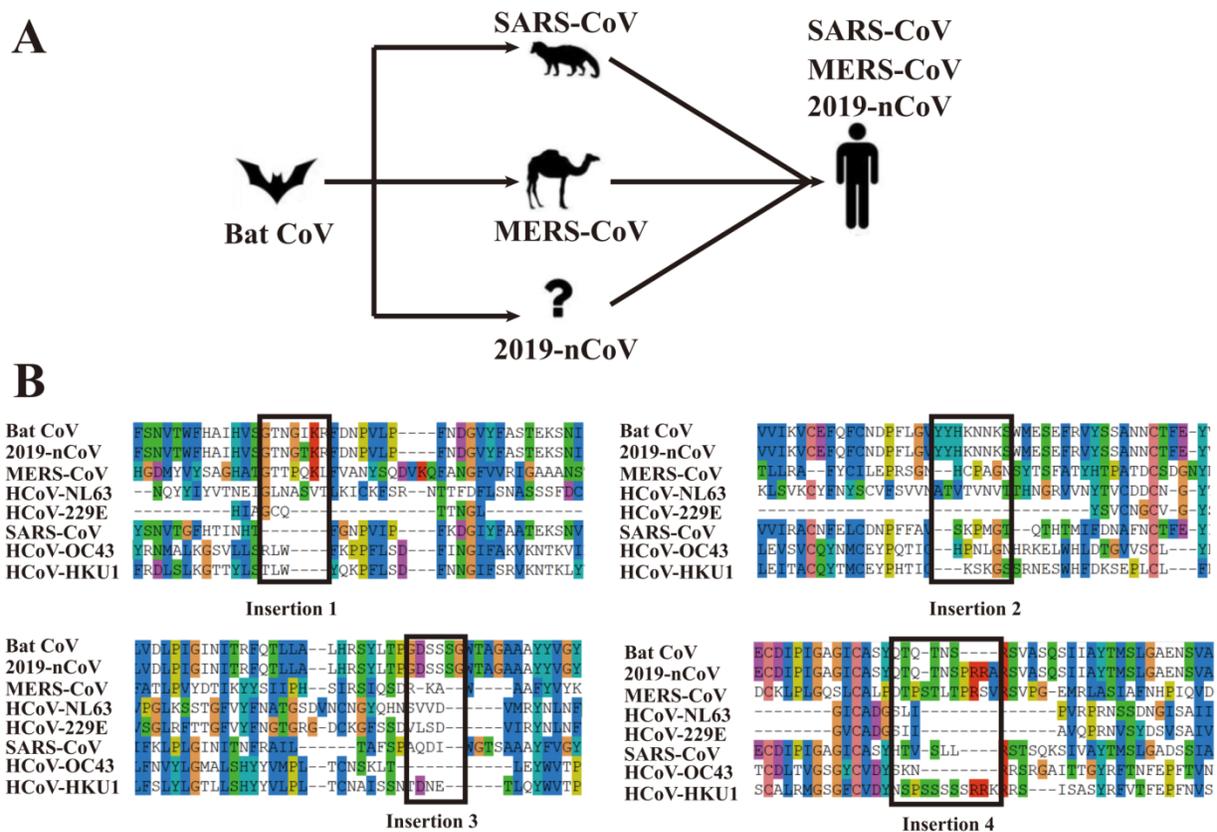

**Figure 3**. Proposed transmission history of representative coronaviruses **(A)** and multiple sequence alignment for the bat coronavirus and 7 known human coronaviruses **(B)**. Bat CoV (QHR63300.1), 2019-nCoV

(QHD43416.1), MERS-CoV (YP_009047204.1), HCoV-NL63 (YP_003767.1), HCoV-229E (NP_073551.1), SARS-CoV (P59594), HCoV-OC43 YP_009555241.1) and HCoV-HKU1 (YP_173238.1) are downloaded from the NCBI and UniProt databases. For brevity, only the regions near the four "insertions" are displayed in the figure.

**Codon Usage Analysis does not Suggest Snakes as Potential Hosts of 2019-nCoV**

Another early study attempting to understanding the infection of 2019-nCoV was performed by Ji *et al.* [4]. In this study, the authors analyzed the Relative Synonymous Codon Usage (RSCU) of 2019-nCoV and 8 vertebrates, including two species of snakes (*Bungarus multicinctus* and *Naja atra*), hedgehog (*Erinaceus europaeus*), bat (*Rhinolophus sinicus*), marmot (*Marmota*), pangolin (*manis javanica*), chicken (*Gallus gallus*), and human (*Homo sapiens*). Among these vertebrates, snakes have the smallest codon usage difference (squared Euclidean distance of RSCU) from 2019-nCoV and were therefore proposed by Ji *et al.* as the intermediate hosts of 2019-nCoV. This conclusion is controversial among virologists, mainly due to the lack of prior biological evidence that coronavirus can infect animals other than mammals and birds [15]. In this report, we further scrutinize the bioinformatics approach and the underlying biological assumptions by a large-scale replication of the RSCU analysis.

The bioinformatics analysis performed in Ji *et al.* study has several limitations. First, there are only approximately 300 protein coding sequences (CDSs) in the NCBI GenBank for the snake species they chose from *Bungarus multicinctus* and *Naja atra*. These CDSs represent <2% of all protein coding genes in a typical snake genome; the genome of King cobra (*Naja hannah*), for example, encodes 18,387 proteins according to UniProt (https://www.uniprot.org/proteomes/UP000018936). The limited numbers of known CDSs in *Bungarus multicinctus* and *Naja atra* mean that the RSCU statistics may not reflect the actual RSCU distribution in the whole genome. Second, the Codon Usage Database used in the analysis of Ji *et al.* has not been updated since 2007; a re-analysis using more recent codon usage database such as CoCoPUTs is needed. Third, only 8 vertebrates were analyzed in their study, while there are >100,000 vertebrates with at least one CDS in the NCBI GenBank database. Finally, there is no established evidence that viruses evolve their codon usage to resemble that of their animal hosts [16]; this calls for a careful benchmark of RSCU analysis in terms of its ability to re-discover known hosts of characterized viruses.

To address these issues, we re-implemented the RSCU comparison algorithm proposed by Ji *et al.* to analyze the codon usage in the 2019-nCoV genome (NCBI accession MN908947.3) and those of all 102,367 vertebrate species in the CoCoPUTS database. To test whether this kind of analysis can recover known hosts of well-studied coronavirus, severe acute respiratory syndrome-related coronavirus (SARS-CoV, NCBI accession NC_004718) and Middle East respiratory syndrome-related coronavirus (MERS-CoV, NCBI accession NC_019843) are also included. Codon usage frequency are converted to squared Euclidean distance of RSCU in two separate analyses: one based on all vertebrates (**Figure 4A-C**) and the other on the subset of vertebrates with enough statistics, i.e. >2000 known CDSs (**Figure 4D-F**), roughly corresponding to 10% of all protein coding genes in a typical vertebrate genome.

As shown in **Figure 4**, the top-ranking vertebrates with lowest RSCU distances to the three different coronavirus are very similar. For example, for all three viruses, two kinds of frogs (*Megophrys feae* and *Liophryne schlaginhaufeni*) have the smallest RSCU distances among all vertebrates, while another two kinds of frogs (*Xenopus laevis* and *Xenopus tropicalis*) have the smallest RSCU distances among all vertebrates with sufficient sequences. In fact, the RSCU distances between coronaviruses and frogs are consistently smaller than that between coronaviruses and the two snakes (*Bungarus multicinctus* and *Naja atra*) analyzed in Ji *et al.*, as shown in **Table 5**.

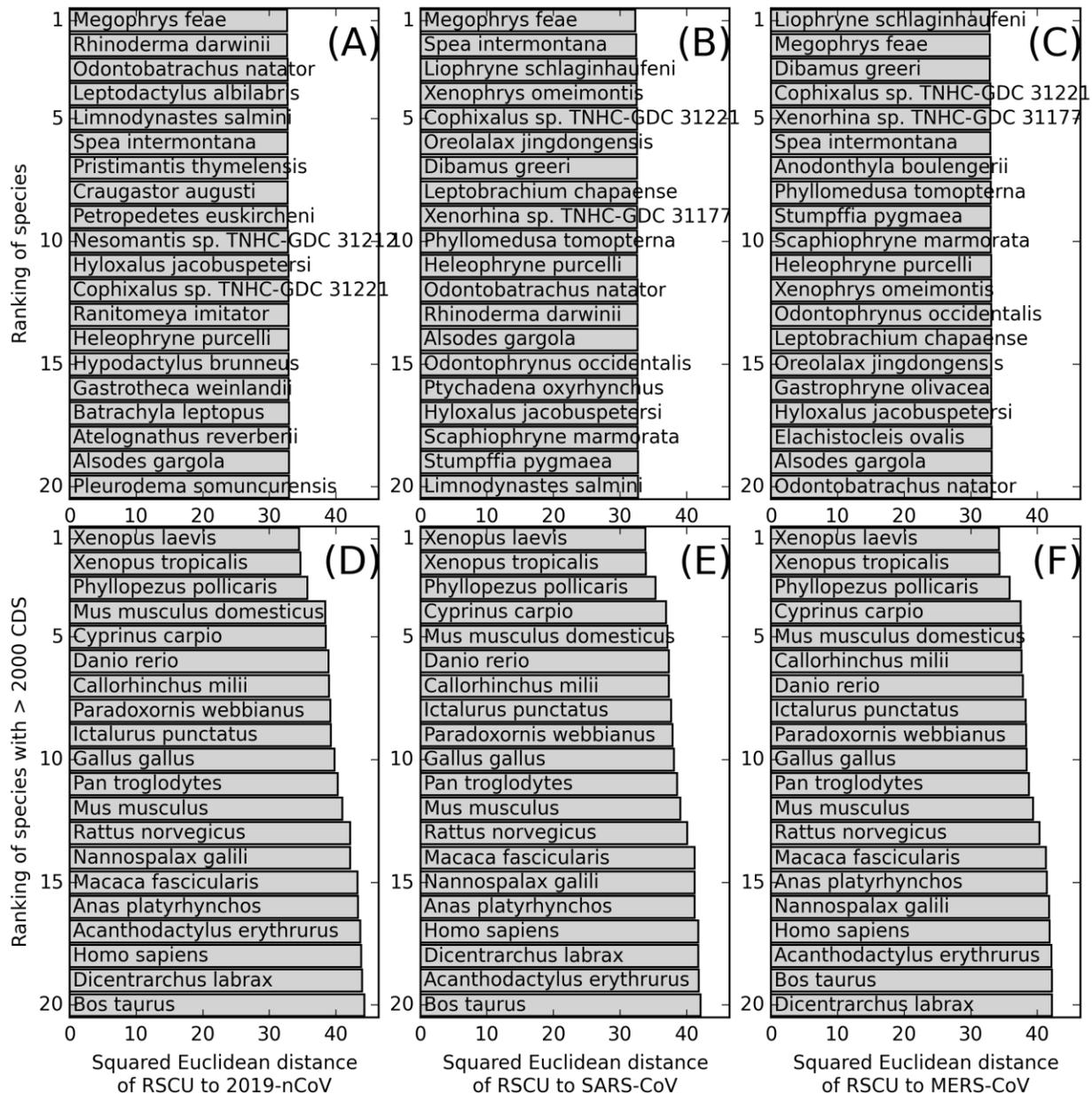

**Figure 4.** Top 20 vertebrate species ranked in ascending order of squared Euclidean distance of RSCU to 2019-nCoV **(A, D)**, SARS-CoV **(B, E)**, and to MERS-CoV **(C, F)**. The upper panels **(A-C)** are for all vertebrates with known codon usage in CoCoPUTS database, while the lower panel **(D-F)** is for the subset of species where more than 2000 CDSs are available for codon usage calculation.

**Table 5** also showed that the RSCU analysis cannot identify the known natural, intermediate, and eventual hosts (*Rhinolophus sinicus*, *Paguma larvata*, and *Homo sapiens*) of SARS-CoV or the hosts (*Camelus dromedarius* and *Homo sapiens*) of MERS-CoV, as the RSCU distances of these vertebrates are all larger than the cold blooded vertebrates (frogs and snakes). In fact,

SARS-CoV and MERS-CoV, the two coronaviruses known to utilize different intermediate hosts (*Paguma larvata* and *Camelus dromedarius*), have almost no difference in RSCU. These data suggest that the RSCU analysis proposed by Ji *et al*. cannot provide conclusive evidence for the natural, intermediate or eventual hosts of coronavirus. Based on our current knowledge of existing zoonotic coronaviruses [15], we conceive that, if 2019-nCoV requires an intermediate host, this host is more likely to be a warm-blooded vertebrate (a mammal or a bird), than snakes.

**Table 5.** Squared Euclidean distances among coronaviruses and representative vertebrates.

| Species | 2019-nCoV | SARS-CoV | MERS-CoV | Comment |
| --- | --- | --- | --- | --- |
| 2019-nCoV | 0 | 0.28 | 0.40 | Coronaviruses |
| SARS-CoV | 0.28 | 0 | 0.12 | |
| MERS-CoV | 0.40 | 0.12 | 0 | |
| *Megophrys feae* | 32.64 | 32.28 | 32.86 | Vertebrates with the closest RSCU to the coronaviruses |
| *Liophryne schlaginhaufeni* | 33.07 | 32.50 | 32.80 | |
| *Rhinolophus affinis* | 37.58 | 36.47 | 36.50 | Natural host of 2019-nCoV |
| *Rhinolophus sinicus* | 37.55 | 36.25 | 36.49 | Natural host of SARS-CoV |
| *Bungarus multicinctus* | 34.48 | 34.36 | 34.84 | Proposed intermediate hosts of 2019-nCoV by Ji *et al*. |
| *Naja atra* | 34.44 | 33.98 | 34.39 | |
| *Paguma larvata* | 47.34 | 45.08 | 45.04 | Intermediate host of 2019-nCoV |
| *Camelus dromedarius* | 46.47 | 44.28 | 44.63 | Intermediate host of MERS-CoV |
| *Homo sapiens* | 43.81 | 41.71 | 41.84 | Eventual host |

## Discussion and Conclusion

Due to the scarcity of experimental and clinical data, as well as the urgency to understand the infectivity of the deadly coronaviruses, we have been increasingly relying on computational analyses to study the 2019-nCoV virus in terms of protein structures, functions, phylogeny, and interactions at both molecular and organismal levels. Indeed, within less than a month of the publication of the 2019-nCoV genome in January 2020, multiple bioinformatics analyses regarding 2019-nCoV have been either published or posted as preprint. While such expeditious analyses provide much needed insights into the biology of the 2019-nCoV virus, there is a caution to avoid over-interpretation of the data at the absence of comprehensive benchmarks or follow-up experimental validations. In this report, we have investigated two recently published computational analyses regarding intermediate host identification and the analysis of spike protein insertions. In both cases, we found that the conclusions proposed by the original studies do not hold in the face of more comprehensive replications of these analyses. We hope our analysis presented herein can clarify some of the misinterpretations resulted from previous bioinformatics analyses for the 2019-nCoV virus.

## Methods

### Protein Sequence Alignment

Global protein sequence alignment of the full-length coronavirus spike proteins was performed by MUSCLE [17] and visualized by SeaView [18].

**Structure Prediction of Spike-ACE2 complex**

We use C-I-TASSER [5] to create structural models of the full-length spike protein. Here, C-I-TASSER is an extended pipeline of I-TASSER[19] and utilizes the deep convolutional neural-network based contact-maps [20] to guide the Monte Carlo fragment assembly simulations. Since the RBD domain of spike exhibits different conformations relative to the remaining portion of the protein, the DEMO pipeline [8] was then used to re-assembly the domains and to construct a complex structure consisting of spike trimer and the extracellular domain of human ACE2, using the ACE2-bound conformation 2 of SARS-CoV spike glycoprotein (PDB ID: 6ACJ) as a template. Our complex modeling did not use the template originally used in the Pradhan *et al*. study (PDB ID: 6ACD) because it did not include the ACE2 receptor.

**Relative Synonymous Codon Usage (RSCU) Analysis**

As per the previous study [4], the RSCU for codon *j* in a species is calculated as

$$X_j = p_j \cdot k_j \tag{1}$$

where $k_j$ is the number of codons synonymous to codon *j* (including *j* itself), $p_j$ is the probability of the respective amino acid being encoded by codon *j* among all $k_j$ synonymous codons in the CDSs of the whole genome. The difference of codon usage in two different species (a virus versus a vertebrate in our case) is defined by squared Euclidean distance of RSCU, i.e.,

$$d = \sum_{j=1}^{N} (X_j - X'_j)^2 \tag{2}$$

Here, *N*=61 is the number of codons that encodes amino acids, thereby excluding the 3 stop codons. $X_j$ and $X_j'$ are the RSCU for codon *j* in the virus and in the vertebrate, respectively. In our report, the codon usages of all vertebrates are taken from the CoCoPUTS [21] database, which was last updated in January 2020. This database was therefore much more recent than the Codon Usage Database [22] last updated in 2007 that was used in the previous research [4]. To obtain the codon usage of coronaviruses, we imported the GenBank annotations of the three coronavirus genomes to SnapGene (GSL Biotech LLC] to export the codon usage table based on the GenBank annotations. CodonW [23] was not used for codon usage calculation as in the previous study because it cannot account for the -1 frameshift translation of the first open reading frame (ORF) in the coronavirus genome.

## Acknowledgements

This work is supported in part by the National Institute of General Medical Sciences (GM083107, GM116960), the National Institute of Allergy and Infectious Diseases (AI134678), and the National Science Foundation (DBI1564756, IIS1901191). This work used the Extreme Science and Engineering Discovery Environment (XSEDE),[24] which is supported by National Science Foundation (ACI-1548562).

## References

1. Zhou, P.; Yang, X.-L.; Wang, X.-G.; Hu, B.; Zhang, L.; Zhang, W.; Si, H.-R.; Zhu, Y.; Li, B.; Huang, C.-L.; Chen, H.-D.; Chen, J.; Luo, Y.; Guo, H.; Jiang, R.-D.; Liu, M.-Q.; Chen, Y.; Shen, X.-R.; Wang, X.; Zheng, X.-S.; Zhao, K.; Chen, Q.-J.; Deng, F.; Liu, L.-L.; Yan, B.; Zhan, F.-X.; Wang, Y.-Y.; Xiao, G.-F.; Shi, Z.-L., A pneumonia outbreak associated with a new coronavirus of probable bat origin. *Nature* **2020**.
2. Letko, M.; Munster, V., Functional assessment of cell entry and receptor usage for lineage B β-coronaviruses, including 2019-nCoV. *bioRxiv* **2020**, 2020.01.22.915660.
3. Pradhan, P.; Pandey, A. K.; Mishra, A.; Gupta, P.; Tripathi, P. K.; Menon, M. B.; Gomes, J.; Vivekanandan, P.; Kundu, B., Uncanny similarity of unique inserts in the 2019-nCoV spike protein to HIV-1 gp120 and Gag. *bioRxiv* **2020**, 2020.01.30.927871.
4. Ji, W.; Wang, W.; Zhao, X.; Zai, J.; Li, X., Homologous recombination within the spike glycoprotein of the newly identified coronavirus may boost cross-species transmission from snake to human. *Journal of Medical Virology* **2020**.
5. Zheng, W.; Li, Y.; Zhang, C. X.; Pearce, R.; Mortuza, S. M.; Zhang, Y., Deep-learning contact-map guided protein structure prediction in CASP13. *Proteins* **2019**.
6. Zhang, Y.; Skolnick, J., Scoring function for automated assessment of protein structure template quality. *Proteins* **2004,** *57* (4), 702-710.
7. Song, W.; Gui, M.; Wang, X.; Xiang, Y., Cryo-EM structure of the SARS coronavirus spike glycoprotein in complex with its host cell receptor ACE2. *PLoS Pathog* **2018,** *14* (8), e1007236.
8. Zhou, X. G.; Hu, J.; Zhang, C. X.; Zhang, G. J.; Zhang, Y., Assembling multidomain protein structures through analogous global structural alignments. *Proceedings of the National Academy of Sciences of the United States of America* **2019,** *116* (32), 15930-15938.
9. Altschul, S. F.; Madden, T. L.; Schäffer, A. A.; Zhang, J.; Zhang, Z.; Miller, W.; Lipman, D. J., Gapped BLAST and PSI-BLAST: a new generation of protein database search programs. *Nucleic Acids Research* **1997,** *25* (17), 3389-3402.
10. Li, W.; Shi, Z.; Yu, M.; Ren, W.; Smith, C.; Epstein, J. H.; Wang, H.; Crameri, G.; Hu, Z.; Zhang, H.; Zhang, J.; McEachern, J.; Field, H.; Daszak, P.; Eaton, B. T.; Zhang, S.; Wang, L.-F., Bats Are Natural Reservoirs of SARS-Like Coronaviruses. *Science* **2005,** *310* (5748), 676.
11. Wang, Q.; Qi, J.; Yuan, Y.; Xuan, Y.; Han, P.; Wan, Y.; Ji, W.; Li, Y.; Wu, Y.; Wang, J.; Iwamoto, A.; Woo, Patrick C. Y.; Yuen, K.-Y.; Yan, J.; Lu, G.; Gao, George F., Bat Origins of MERS-CoV Supported by Bat Coronavirus HKU4 Usage of Human Receptor CD26. *Cell Host & Microbe* **2014,** *16* (3), 328-337.
12. Corman, V. M.; Baldwin, H. J.; Tateno, A. F.; Zerbinati, R. M.; Annan, A.; Owusu, M.; Nkrumah, E. E.; Maganga, G. D.; Oppong, S.; Adu-Sarkodie, Y.; Vallo, P.; da Silva Filho, L. V. R. F.; Leroy, E. M.; Thiel, V.; van der Hoek, L.; Poon, L. L. M.; Tschapka, M.; Drosten, C.; Drexler, J. F., Evidence for an Ancestral Association of Human Coronavirus 229E with Bats. *Journal of Virology* **2015,** *89* (23), 11858.
13. Hu, B.; Zeng, L.-P.; Yang, X.-L.; Ge, X.-Y.; Zhang, W.; Li, B.; Xie, J.-Z.; Shen, X.-R.; Zhang, Y.-Z.; Wang, N.; Luo, D.-S.; Zheng, X.-S.; Wang, M.-N.; Daszak, P.; Wang, L.-F.; Cui, J.; Shi, Z.-L., Discovery of a rich gene pool of bat SARS-related coronaviruses provides new insights into the origin of SARS coronavirus. *PLOS Pathogens* **2017,** *13* (11), e1006698.
14. GISAID, Phylogenetic tree of CoV families (larger context, including nearest bat precursor) (2020-02-03). *https://platform.gisaid.org/epi3/frontend#lightbox983930542* **2020**.
15. Callaway, E.; Cyranoski, D., Why snakes probably aren't spreading the new China virus. *Nature* **2020**.
16. Meintjes, P. L.; Rodrigo, A. G., Evolution of relative synonymous codon usage in Human Immunodeficiency Virus type-1. *Journal of bioinformatics and computational biology* **2005,** *3* (01), 157-168.
17. Edgar, R. C., MUSCLE: multiple sequence alignment with high accuracy and high throughput. *Nucleic Acids Research* **2004,** *32* (5), 1792-1797.

18. Gouy, M.; Guindon, S.; Gascuel, O., SeaView Version 4: A Multiplatform Graphical User Interface for Sequence Alignment and Phylogenetic Tree Building. *Molecular Biology and Evolution* **2009,** *27* (2), 221-224.
19. Yang, J.; Yan, R.; Roy, A.; Xu, D.; Poisson, J.; Zhang, Y., The I-TASSER Suite: protein structure and function prediction. *Nature Methods* **2015,** *12* (1), 7-8.
20. Li, Y.; Zhang, C.; Bell, E. W.; Yu, D. J.; Zhang, Y., Ensembling multiple raw coevolutionary features with deep residual neural networks for contact-map prediction in CASP13. *Proteins* **2019**.
21. Alexaki, A.; Kames, J.; Holcomb, D. D.; Athey, J.; Santana-Quintero, L. V.; Lam, P. V. N.; Hamasaki-Katagiri, N.; Osipova, E.; Simonyan, V.; Bar, H., Codon and Codon-Pair Usage Tables (CoCoPUTs): facilitating genetic variation analyses and recombinant gene design. *Journal of molecular biology* **2019,** *431* (13), 2434-2441.
22. Nakamura, Y.; Gojobori, T.; Ikemura, T., Codon usage tabulated from international DNA sequence databases: status for the year 2000. *Nucleic Acids Research* **2000,** *28* (1), 292-292.
23. Peden, J. F., Analysis of codon usage. **2000**.
24. Towns, J.; Cockerill, T.; Dahan, M.; Foster, I.; Gaither, K.; Grimshaw, A.; Hazlewood, V.; Lathrop, S.; Lifka, D.; Peterson, G. D.; Roskies, R.; Scott, J. R.; Wilkins-Diehr, N., XSEDE: Accelerating Scientific Discovery. *Comput Sci Eng* **2014,** *16* (5), 62-74.